# Hidden orbital polarization in diamond, silicon, germanium, gallium arsenide and layered materials

*Ji Hoon Ryoo and Cheol-Hwan Park\**

Department of Physics, Seoul National University, Seoul 08826, Korea

\* Correspondence: Prof. Cheol-Hwan Park
Postal address: Department of Physics, Seoul National University, 1 Gwanak-ro, Gwanak-gu, Seoul 08826, Korea
Telephone: +8228804396
E-mail: cheolhwan@snu.ac.kr

**ABSTRACT**

**It was previously believed that the Bloch electronic states of non-magnetic materials with inversion symmetry cannot have finite spin polarizations. However, since the seminal work by Zhang *et al*. [*Nat. Phys.* 10, 387-393 (2014)] on local spin polarizations of Bloch states in non-magnetic, centrosymmetric materials, the scope of spintronics has been significantly broadened. Here, we show, using a framework that is universally applicable independent of whether hidden spin polarizations are small (e.g., diamond, Si, Ge, and GaAs) or large (e.g., $MoS_2$ and $WSe_2$), that the corresponding quantity arising from orbital—instead of spin—degrees of freedom, the *hidden orbital polarization*, is (i) much more abundant in nature since it exists even without spin-orbit coupling and (ii) more fundamental since the interband matrix elements of the site-dependent orbital angular momentum operator determines the hidden spin polarization. We predict that the**

**hidden spin polarization of transition metal dichalcogenides is reduced significantly upon compression. We suggest experimental signatures of hidden orbital polarization from photoemission spectroscopies and demonstrate that the current-induced hidden orbital polarization may play a far more important role than its spin counterpart in antiferromagnetic information technology by calculating the current-driven antiferromagnetism in compressed silicon.**

## INTRODUCTION

Electronic states at a given Bloch wavevector in non-magnetic materials with inversion symmetry are degenerate. Until recently, it was believed that there is no spatial spin distribution if averaged over these two spin-degenerate states. However, it has been found that even in centrosymmetric, non-magnetic crystals, the degenerate Bloch states can have local spin polarization if atoms are not at an inversion center.[1] Reference 1 reported that the lack of the local inversion symmetry at atomic sites leads to hidden, or site-dependent, spin polarization, expanding the scope of spintronics significantly, even to bulk materials with global inversion symmetry.

On the other hand, the orbital contribution to the magnetic moment of solids can be sizable (e.g., References 2 and 3) and even larger than the spin contribution.[4] The orbital magnetization becomes more important than the spin magnetization in some physical phenomena, e.g., current-induced magnetization[5] and the gyrotropic magnetic effect,[6] if the spin-orbit coupling (SOC) is weak. Additionally, the important role of orbital polarization in Rashba-split bands[7-9] and quantum anomalous Hall phases[10] of systems without inversion symmetry has been explored.

In this paper, we report the finding that the hidden, or sublattice-dependent, orbital polarization of Bloch states of centrosymmetric materials can be large (on the order of $\hbar$) even without SOC by using the simplest, best-known materials, such as diamond, Si, and Ge, as examples. We describe that, in any non-magnetic, centrosymmetric material, including the aforementioned zinc-blende materials and layered materials such as $MoS_2$ and $WSe_2$, in which the hidden spin polarization is quite large, the hidden spin polarization is completely determined by the interband matrix elements of the site-dependent orbital angular momentum operator. This finding, together with the fact that in materials with weak SOC the hidden spin polarization is small or absent, suggests that the hidden orbital polarization is a more fundamental quantity. We show that the sublattice-dependent spin-orbital texture of centrosymmetric crystals is qualitatively different from that of non-centrosymmetric crystals and that the hidden orbital polarization can play an important role in current-induced magnetization[5] of both centrosymmetric materials and non-centrosymmetric materials such as GaAs. We then discuss the experimental evidence from photoemission spectroscopies and the technological implications of our findings in antiferromagnetic information technology using current-induced hidden orbital polarizations, which, according to our calculations, could be much more important than their spin counterpart.

## MATERIALS AND METHODS

We calculated the electronic structures of diamond, Si, Ge and GaAs by using a tight-binding model including atomic $s$ and $p$ orbitals ($sp^3s^*$ model)[11] and the on-site spin-orbit coupling term $\Delta H_{\text{SOC}} = \left(\alpha^A \mathbf{L}^A \cdot \mathbf{S} + \alpha^{\bar{A}} \mathbf{L}^{\bar{A}} \cdot \mathbf{S}\right)/\hbar^2$, in which $A$ and $\bar{A}$ denote the two sublattices in the zinc-blende structure (see Figure 1b), $\alpha^A$ and $\alpha^{\bar{A}}$ are atomic spin-orbit coupling strengths, and the local orbital angular momentum operator $\mathbf{L}^\beta$ ($\beta = A, \bar{A}$) for each sublattice is defined as $L_i^\beta = -i\hbar \sum_{j,k} \epsilon_{ijk} |p_j, \beta\rangle\langle p_k, \beta|$, where $\epsilon_{ijk}$ is the Levi-Civita symbol and $|p_j, \beta\rangle$ is the Bloch sum of $p_j$ orbitals at sublattice $\beta$. This type of model[12] has been used in studies of Rashba splitting and spin-orbital textures.[7,13,14]

**Figure 1.** The hidden orbital polarization in diamond without spin-orbit coupling. (a) The electronic band structure of diamond. $n_v^{\max}$ is the band index of the highest-energy valence band. (b) The local orbital polarization of state P shown in (a). (c-f) The local orbital texture at the $A$ sublattice of diamond on the $k_z = 0$ plane [(c-d)] and on the $k_x + k_y + k_z = 0$ plane [(e-f)]. The $x'$, $y'$, and $z'$ axes point in the $[1\bar{1}0]$, $[11\bar{2}]$, and $[111]$ directions, respectively.

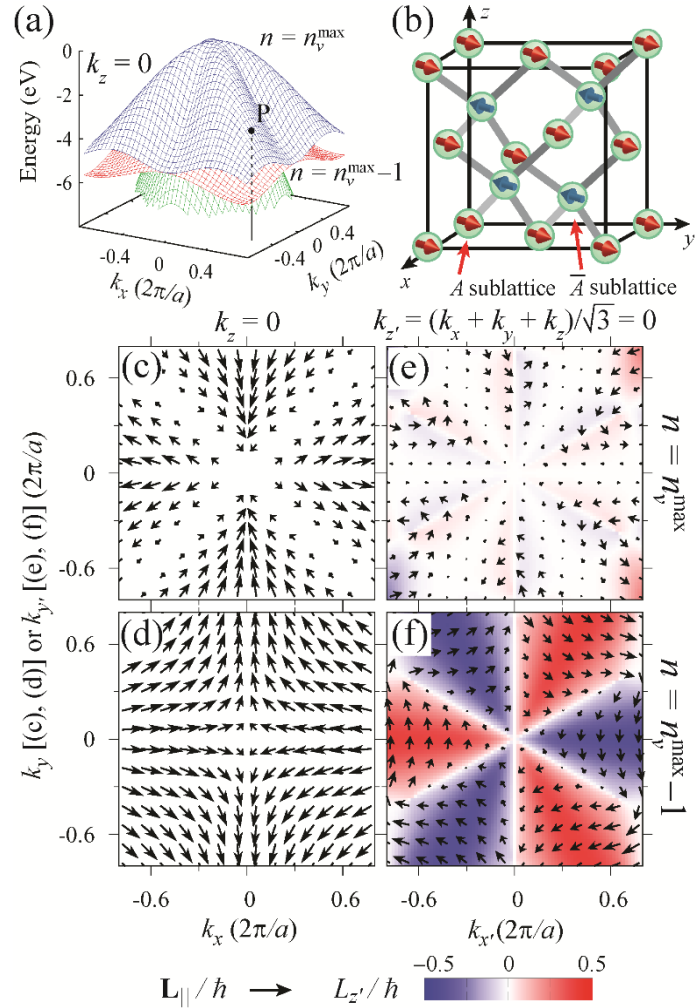

## RESULTS AND DISCUSSION

First, we discuss the orbital polarization of diamond, whose SOC is negligible. If SOC is neglected, the spin-up and -down states have the same energy and orbital wavefunction. Figure 1b shows the local orbital polarization $\langle L^\beta \rangle_{n\mathbf{k}} = \langle n\mathbf{k} | L^\beta | n\mathbf{k} \rangle$ for the orbital part of a Bloch state,

$|n\mathbf{k}\rangle$, corresponding to P in Figure 1a. (Note that the expectation values of $\mathbf{L}^\beta$ with respect to spin-degenerate states $|n\mathbf{k}\rangle \otimes |\uparrow\rangle$ and $|n\mathbf{k}\rangle \otimes |\downarrow\rangle$ are the same.) Since the product of the inversion operator (*P*) and the time reversal operator (*T*) conserves the crystal momentum **k**, all the Bloch states are invariant under *PT* operation if we neglect spin. *A* and $\bar{A}$ are exchanged by *P*. Therefore, the local orbital polarizations at *A* and $\bar{A}$ are of the same magnitude and are anti-parallel to each other; i.e., if we define $\mathbf{L}^{\text{tot}} = \mathbf{L}^A + \mathbf{L}^{\bar{A}}$, $\langle \mathbf{L}^{\text{tot}} \rangle_{n\mathbf{k}} = 0$.

Figures 1c-1f show $\langle \mathbf{L}^A \rangle_{n\mathbf{k}}$ for the (spin-degenerate) highest-energy valence bands and the second-highest-energy valence bands of diamond, which is on the order of $\hbar$ except on some symmetry lines. Therefore, the orbital polarization on each sublattice of a centrosymmetric material can be large. In the $k_z = 0$ plane, since **k** is invariant under $C_2 T$, where $C_2$ is the operator for 180° rotation with respect to the *z*-axis, $\langle \mathbf{L}^A \rangle_{n\mathbf{k}}$ lies in the *xy* plane. We also verified that Si and Ge have similar hidden orbital polarization textures (Supplementary Discussion 1).

Interestingly, the hidden orbital polarization can be large even when the total orbital angular momentum is quenched. Note that the total orbital angular momentum is a ground-state property of a crystal, whereas the hidden orbital polarization is a property of quasi-particle excitations and is a function of the Bloch wavevector and the band index. Even in a material where *d* orbitals of a transition metal element experience a strong octahedral crystal field and the (total) orbital angular momentum is quenched, for example, when the $t_{2g}$ bands are empty / half occupied / fully occupied, a quasi-particle state (either an electron or a hole) from the $t_{2g}$ bands can still have a large hidden orbital polarization.

Next, we show that in non-magnetic, centrosymmetric materials, the hidden spin polarization is a physical quantity completely determined by the site-dependent orbital angular momentum. When SOC is absent, it is apparent that a hidden spin texture cannot exist in these materials; since the electron potential does not depend on the spin, all bands are spin-degenerate, and each Bloch state cannot have a spatially inhomogeneous spin distribution. Conversely, we showed that there can be a large hidden orbital polarization even when SOC is absent. When there is SOC, the spin-up and spin-down bands mix with each other, but they remain degenerate due to *PT* symmetry. We define the spin or orbital polarization of each band as the average of the expectation values of the two degenerate states.[1]

Let $|n\mathbf{k}s\rangle = |n\mathbf{k}\rangle \otimes |s\rangle$ be spin-degenerate eigenstates of the Hamiltonian without SOC, where $|n\mathbf{k}\rangle$ is the orbital part and $|s\rangle$ is the spin part. In our model, in which SOC is taken into account by $\Delta H_{\text{SOC}} = \alpha(\mathbf{L}^A \cdot \mathbf{S} + \mathbf{L}^{\bar{A}} \cdot \mathbf{S})/\hbar^2$, we can express the local spin polarization $\langle \mathbf{S}^A \rangle_{n\mathbf{k}}^{\text{avg}} = -\langle \mathbf{S}^{\bar{A}} \rangle_{n\mathbf{k}}^{\text{avg}}$ in terms of the matrix element of the site-dependent orbital angular momentum operator using first order perturbation theory:

$$\langle \mathbf{S}^\beta \rangle_{n\mathbf{k}}^{\text{avg}} = \sum_{\substack{m \neq n \\ s,s'}} \frac{\langle n\mathbf{k}s | P^\beta \frac{\hbar \sigma}{2} | m\mathbf{k}s' \rangle \langle m\mathbf{k}s' | \Delta H_{\text{SOC}} | n\mathbf{k}s \rangle + c.c.}{2(E_{n\mathbf{k}} - E_{m\mathbf{k}})}$$

$$= \frac{\alpha}{4} \sum_{m \neq n} \frac{\langle n\mathbf{k} | P^\beta | m\mathbf{k} \rangle \langle m\mathbf{k} | (\mathbf{L}^A + \mathbf{L}^{\bar{A}}) | n\mathbf{k} \rangle + c.c.}{E_{n\mathbf{k}} - E_{m\mathbf{k}}}$$

$$= \frac{\alpha}{2} \sum_{m \neq n} \frac{\langle n\mathbf{k} | P^\beta | m\mathbf{k} \rangle \langle m\mathbf{k} | \mathbf{L}^\beta | n\mathbf{k} \rangle + c.c.}{E_{n\mathbf{k}} - E_{m\mathbf{k}}}. \quad (1)$$

Here, $P^\beta$ is the projection operator onto sublattice $\beta$, $\boldsymbol{\sigma}$ is the Pauli spin matrix, and $E_{n\mathbf{k}}$ is the energy of the state $|n\mathbf{k}\rangle$ when SOC is absent. In the third equality of Equation 1, we have used $[\langle n\mathbf{k}|P^A|m\mathbf{k}\rangle\langle m\mathbf{k}|\mathbf{L}^{\bar{A}}|n\mathbf{k}\rangle]^* = \langle n\mathbf{k}|P^A|m\mathbf{k}\rangle\langle m\mathbf{k}|\mathbf{L}^A|n\mathbf{k}\rangle$, which follows from (i) $(PT)P^A(PT)^{-1} = P^{\bar{A}}$, (ii) $(PT)\mathbf{L}^A(PT)^{-1} = -\mathbf{L}^{\bar{A}}$, (iii) $\langle n\mathbf{k}|P^A|m\mathbf{k}\rangle = -\langle n\mathbf{k}|P^{\bar{A}}|m\mathbf{k}\rangle$ if $n \neq m$, and (iv) $PT|n\mathbf{k}\rangle$ is equal to $|n\mathbf{k}\rangle$ up to a phase factor (recall that $|n\mathbf{k}\rangle$ is the orbital part of the wavefunction). We can then calculate the hidden spin polarization from the site-dependent orbital angular momentum operator using Equation 1, one of our key results.

It is straightforward to extend Equation 1 and calculate higher-order terms in a regime where SOC is not small; even in this regime, the interband matrix elements of the orbital angular momentum operator determine the hidden spin polarization. Additionally, Equation 1 can be easily extended to materials with more than two atoms per unit cell or to cases involving *d* or higher-*l* orbitals.[15]

**Figure 2.** The local spin-orbital texture ($k_z = 0$) at sublattice *A* of the Bloch states of diamond obtained by using $\alpha^C = 4$ meV (the physical value for diamond) and $\alpha^C = 1$ eV.

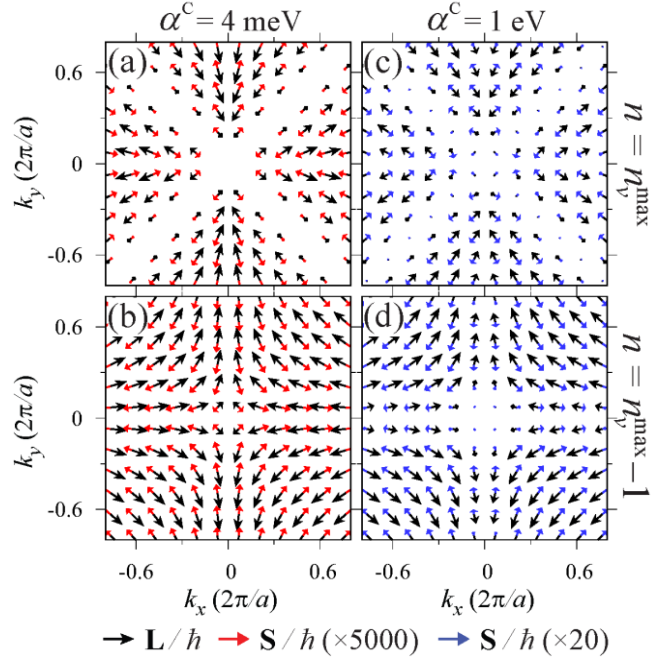

Figure 2 shows the $\langle \mathbf{S}^A \rangle_{n\mathbf{k}}^{\text{avg}}$ of the two highest-energy valence bands of diamond, calculated by direct diagonalization of the Hamiltonian (rather than using Equation 1). Since the SOC in diamond is very weak, $|\langle \mathbf{S}^A \rangle_{n\mathbf{k}}^{\text{avg}}| \ll |\langle \mathbf{L}^A \rangle_{n\mathbf{k}}^{\text{avg}}|$ (Figures 2a and 2b). Quite surprisingly, even if we set the SOC strength $\alpha^C$ to 1 eV, approximately 250 times the physical value, the hidden spin polarization is still an order of magnitude smaller than the orbital polarization (Figures 2c and 2d) because in diamond, the *A* and $\bar{A}$ sublattices are strongly coupled to each other. This fact demonstrates that although the conditions for the existence of hidden orbital and spin polarizations are the same in terms of symmetry, it is more difficult for hidden spin polarizations to be appreciably large (see the analysis on MoS$_2$ and WSe$_2$ below and Supplementary

Discussion 3). However, in some centrosymmetric materials, the hidden spin polarization can be nearly fully polarized;[1] even in this case, our claim that the orbital polarization determines the spin polarization is valid. It is noteworthy that the hidden spin polarization shown in Figure 2 is almost identically reproducible by Equation 1, and the lowest-order result in Equation 1 holds for a wide range of SOC strengths up to $\alpha^C = 1$ eV.

Interestingly, the directions of spin and orbital polarizations are exactly opposite each other (Figure 2). It is difficult to find a simple reason for this (anti-)alignment because Equation 1 expresses the hidden spin polarization in terms of the off-diagonal matrix elements of $\mathbf{L}^A$, rather than the diagonal ones. However, we can understand this behavior in some limited cases (Supplementary Discussion 4).

The hidden spin polarizations in the materials considered above (diamond, silicon and germanium) are much less than 1 %, and even if we hypothetically increase the strength of the spin-orbit coupling of carbon atoms to over 1 eV in our tight-binding model calculations (the physical value of SOC is 4 meV), the hidden spin polarization does not exceed 5 %. In contrast, the hidden spin polarizations in $MoS_2$ and $WSe_2$, whose atomic SOC values are only 0.08 eV and 0.29 eV, respectively, are nearly fully polarized. We investigate this phenomenon and find the origin of such large hidden spin polarizations in $MoS_2$ or $WSe_2$ by extending the analysis on diamond, silicon and germanium. Our tight-binding model is based on Reference 16.

The unit cell of bulk $MoS_2$ consists of two $MoS_2$ units, which are inversion partners [see Figure 3(a)]. The spin (or orbital) polarization of the upper layer in the unit cell points in the opposite direction from that of the lower layer. Figure 3d shows this hidden spin polarization of the lower layer of $MoS_2$ in the highest-energy valence bands at K as a function of the atomic SOC of Mo atoms. The hidden spin polarization is 80 % polarized at the physical value of SOC ($\alpha_{phys}^{Mo} = 0.08$ eV), which is much larger than the hidden spin polarization of diamond (~ 0.01 %).

**Figure 3**. (a) The atomic structure of bulk $MoS_2$. (b) and (c) The electronic band structures of $MoS_2$ obtained by setting the interlayer coupling [(b)] or the atomic SOC [(c)] to zero. (d) The hidden polarizations along $z$ of the lower layer of the highest-energy valence bands at K versus atomic SOC. The fully polarized values are $\langle S^{lower}\rangle_{avg}^{max} = \frac{\hbar}{2}\times 0.5$ and $\langle L^{lower}\rangle_{avg}^{max} = 2\hbar \times 0.5$. The vertical dash-dotted line shows the physical value of the atomic spin-orbit coupling of Mo atoms. (e) The same quantity as in (d), but with the interlayer distance reduced to 85 % of the actual value.

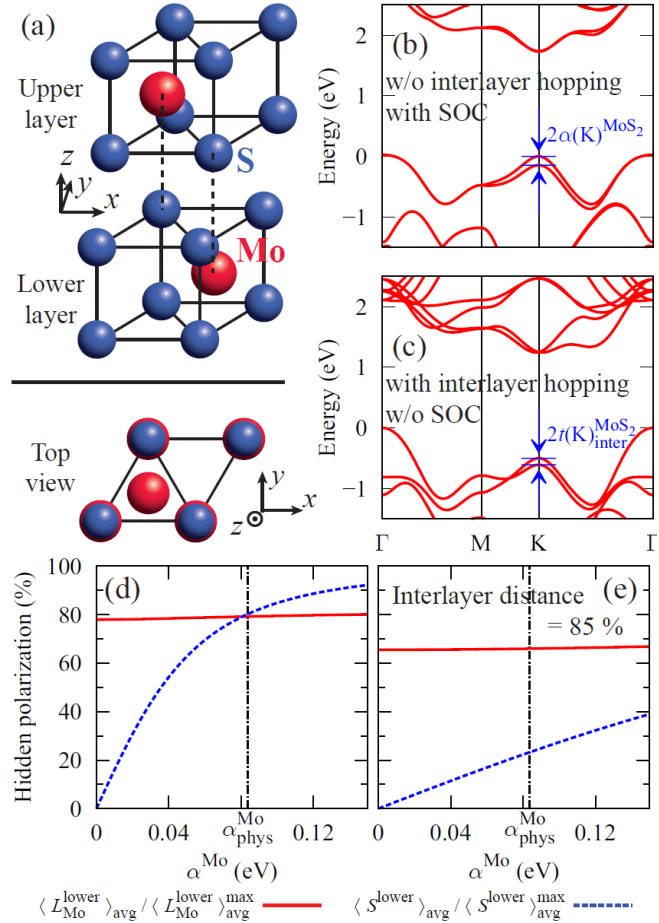

Now, consider the energy splitting among the highest-energy valence states at K due to SOC if there was no interlayer coupling, which we call $2\alpha(K)^{MoS_2}$, and study how the hidden spin polarization varies with $\alpha(K)^{MoS_2}$ (see Figure 3b). For example, $\alpha(K)^{MoS_2} = 0.073$ eV if $\alpha^{Mo} = \alpha_{phys}^{Mo} = 0.084$ eV is used. On the other hand, at zero $\alpha(K)^{MoS_2}$, which is simulated by $\alpha^{Mo} = 0$, there is an energy splitting of 0.11 eV at the top of the valence bands at K among the degenerate doublets due to interlayer splitting (see Figure 3c). Let us call half of this energy splitting $t(K)_{\text{inter}}^{\text{MoS}_2}$ (= 0.053 eV), which vanishes if we set interlayer hopping integrals to zero. Now if $\alpha(K)^{\text{MoS}_2}$, which is proportional to the atomic SOC $\alpha^{\text{Mo}}$, is lower than $t(K)_{\text{inter}}^{\text{MoS}_2}$, the hidden spin polarization is roughly proportional to $\alpha^{\text{Mo}}$, which is consistent with Equation 1 (see Figure 3d). At $\alpha(K)^{\text{MoS}_2}$ values higher than $t(K)_{\text{inter}}^{\text{MoS}_2}$, the wavefunctions of nearby bands are inter-mixed by the SOC, and the hidden spin polarization saturates with $\alpha^{\text{Mo}}$ to 100 % (see Figure 3d). We now can qualitatively understand the results in Figure 3d.

To deepen our understanding of the hidden orbital and spin polarizations in MoS$_2$, we hypothetically decreased the interlayer distance between each MoS$_2$ layer by 15 % and modified the interlayer hopping integrals according to the scheme in Reference 16. The hidden spin polarization of the highest-energy valence bands (doublet) at K is plotted in Figure 3e as a function of $\alpha^{\text{Mo}}$. In this case, the hidden spin polarization is only 20 % of the fully polarized value and scales linearly with $\alpha^{\text{Mo}}$ around $\alpha^{\text{Mo}} = \alpha_{\text{phys}}^{\text{Mo}} = 0.084$ eV. Note that as a byproduct, our calculations give us insight into MoS$_2$ under a high pressure. A 15 % compression of transition metal dichalcogenide compounds has already been achieved in recent high-pressure experiments;[17] the hidden spin polarization of MoS$_2$ under pressure is likely to be significantly lower than that of MoS$_2$ not under pressure. To obtain a quantitative prediction, first-principles calculations with structural optimizations are necessary. Although such first-principles calculations are beyond the scope of this study, the qualitative prediction of the reduction of hidden spin polarizations in inversion-symmetric transition metal dichalcogenides due to pressure remains meaningful.

We now perform a similar analysis on WSe$_2$. In this material, $\alpha(K)_{\text{phys}}^{\text{Wse}_2} = 0.25$ eV (Figure 4a), and $t(K)_{\text{inter}}^{\text{Wse}_2} = 0.036$ eV (Figure 4b): the SOC is stronger in Wse$_2$ than in MoS$_2$, and the interlayer coupling is weaker in WSe$_2$ than in MoS$_2$. Both of these differences lead to stronger hidden spin polarizations in WSe$_2$ than in MoS$_2$, in agreement with the results of our calculations (Figure 4c and 4d).

**Figure 4.** (a) and (b) The electronic band structures of WSe$_2$ obtained by setting the interlayer coupling to zero [(a)] or the atomic SOC [(b)] to zero. (c) The hidden polarizations along $z$ of the lower layer of the highest-energy valence bands at K versus atomic SOC. The fully polarized values are $\langle S^{\text{lower}} \rangle_{\text{avg}}^{\max} = \frac{\hbar}{2} \times 0.5$ and $\langle L^{\text{lower}} \rangle_{\text{avg}}^{\max} = 2\hbar \times 0.5$. The vertical dash-dotted line shows the physical value of the atomic spin-orbit coupling of W atoms. (d) The same quantity as in (d), but with the interlayer distance reduced to 85 % of the actual value.

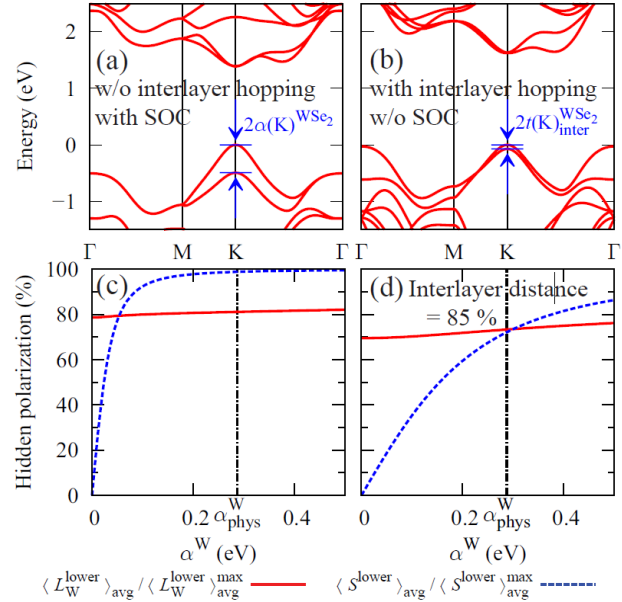

Although Equation 1, which is the result of a perturbation theory calculation, is not applicable *per se* when the SOC is stronger than the interlayer coupling of MoS$_2$ and WSe$_2$, it is still true (as we discussed in the main manuscript) that the site-dependent orbital angular momentum determines the hidden spin polarization. In the case of bulk MoS$_2$ and WSe$_2$, because both the strength of the SOC and the interlayer coupling are smaller than the intralayer coupling, we can determine the hidden spin polarization from the hidden orbital polarization by treating both the strength of the SOC and the interlayer coupling as perturbations:

$$\langle S^{\text{lower}} \rangle_{\text{avg}} = \langle S^{\text{lower}} \rangle_{\text{avg}}^{\max} \times \frac{\alpha^{\text{MoS}_2}(K)}{\left(t_{\text{inter}}^{\text{MoS}_2}(K)^2 + \alpha^{\text{MoS}_2}(K)^2\right)^{1/2}} \qquad (2)$$

as explained in Supplementary Discussion 2. As mentioned previously, this result can also be obtained by using higher-order perturbation theory only with respect to SOC; in hindsight,

expanding the above equation, we know that these higher-order terms should coincide term by term with

$$\langle S^{\text{lower}}\rangle_{\text{avg}} = \langle S^{\text{lower}}\rangle_{\text{avg}}^{\max} \times \left[\frac{\alpha^{\text{MoS}_2}(K)}{t_{\text{inter}}^{\text{MoS}_2}(K)} - \frac{1}{2}\left(\frac{\alpha^{\text{MoS}_2}(K)}{t_{\text{inter}}^{\text{MoS}_2}(K)}\right)^2 + \cdots\right].$$

In contrast with $MoS_2$ or $WSe_2$, in which the two subsystems comprising the unit cell are weakly coupled, the two sublattices of diamond or silicon are strongly coupled to each other; hence, the typical energy separation between energy bands is on the order of the nearest-neighbor hopping integral (a few eVs) and is much larger than the SOC. Remarkably, we can now understand, from the same principles, why the hidden spin polarization in diamond, silicon or germanium is very small and why that in $MoS_2$ or $WSe_2$ is very large.

Now, we turn our attention to the hidden orbital polarization. While the hidden spin polarization depends strongly on the strength of the SOC, the hidden orbital polarization is rather insensitive to it (see Figures 3d, 3e, 4c and 4d). The reason is twofold: (1) the hidden orbital polarization is already large without the SOC, and (2) the two highest-energy valence band doublets at K have approximately the same hidden orbital polarization. This supports our primary claim that hidden orbital polarizations are much more widespread in nature than their spin counterparts. The hidden spin polarization is large only if an inversion center is not located at an atomic site and the SOC is stronger than the energy separation between the bands of interest and other nearby bands. In contrast, the hidden orbital polarization can be large in general if only the first (symmetry-related) condition is met. Our analysis of the connection between the hidden spin and orbital polarizations in $MoS_2$ and $WSe_2$ has not been performed in previous studies, in which the focus has been solely on the hidden spin polarization.

Thus far, we have discussed the hidden spin and orbital polarizations of centrosymmetric materials. The spin texture in non-centrosymmetric materials is qualitatively different from that in centrosymmetric systems. Without SOC, all the electronic energy bands of a non-magnetic material are spin degenerate. Contrary to centrosymmetric systems, in which the SOC intermixes spin-up and spin-down components without lifting the degeneracy, the SOC in non-centrosymmetric systems lifts this degeneracy.

**Figure 5.** The site-dependent spin and orbital polarizations ($k_z = 0$) of the two bands of GaAs that originate from the spin-degenerate, second-highest-energy doublet among valence bands when SOC is neglected. (a) and (b), (c) and (d), and (e) and (f) show the quantities of the upper spin-split band, those of the lower spin-split band, and their averages, respectively.

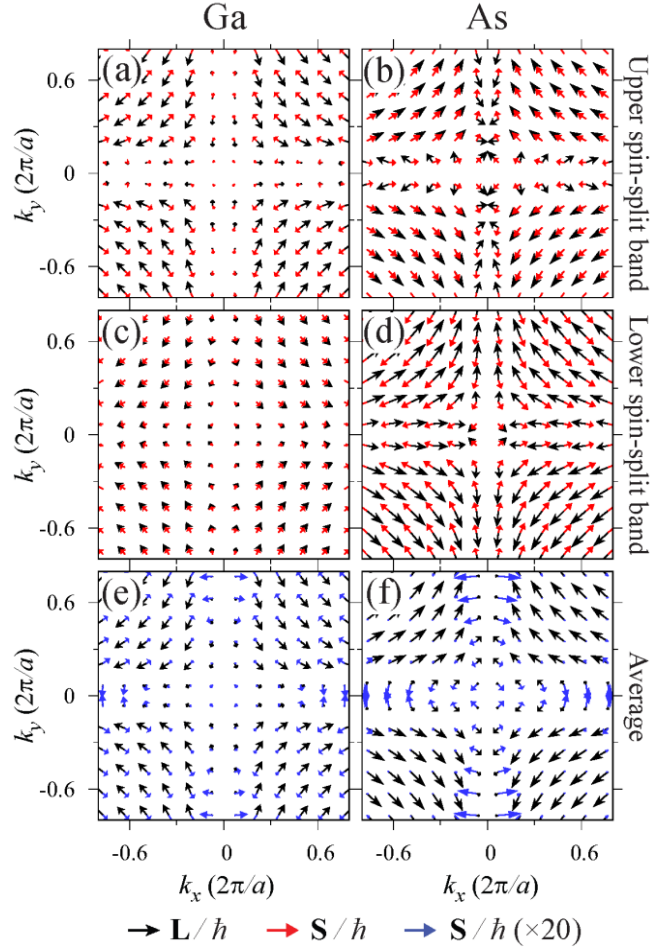

Figures 5a-5d show the spin-orbital texture ($k_z = 0$) of the two bands of GaAs split from the second-highest, spin-degenerate valence bands when SOC is absent. The orbital polarization of As atoms is approximately twice that of Ga atoms, which can be attributed to the lower on-site potential energy of As atoms. Their directions are opposite to each other, similar to the hidden orbital polarizations at the $A$ and $\bar{A}$ sublattices of diamond (Figures 1 and 2).

Comparing the upper spin-split band (Figures 5a-5b) and the lower spin-split band (Figures 5c-5d), we note that, except near the $k_x$ or $k_y$ axes, the orbital polarizations of the upper and lower bands are approximately the same because the SOC mixes only the spin-up and -down bands together; its magnitude is smaller than the energy distance from those bands to other adjacent bands.

The spin texture of GaAs in Figure 5 demonstrates the following features: (i) excluding the regions near $k_x = 0$ or $k_y = 0$ where four bands are degenerate if the SOC is absent, the spin polarization is parallel or anti-parallel to the orbital polarization, (ii) the spin polarization at the Ga atoms and As atoms of each spin-split band are parallel to each other, in contrast with the hidden spin polarization of diamond, i.e., $\langle \mathbf{S}^A \rangle_{n\mathbf{k}}^{\text{avg}}$ and $\langle \mathbf{S}^{\bar{A}} \rangle_{n\mathbf{k}}^{\text{avg}}$ are anti-parallel to each other, and (iii) the spin is almost fully polarized in each band. These observations also hold for other bands of GaAs (Supplementary Discussion 1).

These features can be explained as follows. When the SOC is neglected, the spin-up and -down bands are degenerate and share the common orbital wavefunction $|n\mathbf{k}\rangle$. Within degenerate perturbation theory, the effect of the SOC is described by diagonalizing $\Delta H_{\text{SOC}} = (\alpha^{\text{Ga}} \mathbf{L}^{\text{Ga}} + \alpha^{\text{As}} \mathbf{L}^{\text{As}}) \cdot \mathbf{S}/\hbar^2$ in the two-dimensional Hilbert space spanned by the spin-up and spin-down states. (We set $\alpha^{\text{Ga}} = 0.12$ eV and $\alpha^{\text{As}} = 0.28$ eV.[18]) Therefore, if there is no other degeneracy, the direction of the spin polarization of one spin-split band is parallel to $\langle n\mathbf{k}|[\alpha^{\text{Ga}} \mathbf{L}^{\text{Ga}} + \alpha^{\text{As}} \mathbf{L}^{\text{As}}]|n\mathbf{k}\rangle$ (we will denote a unit vector aligned in this direction as $\hat{\rho}_{n\mathbf{k}}$) and the spin polarization of the other spin-split band points in the opposite direction. We define $|\uparrow; \hat{\rho}_{n\mathbf{k}}\rangle$ and $|\downarrow; \hat{\rho}_{n\mathbf{k}}\rangle$ as the spinors whose spin quantization axes are parallel to and anti-parallel to $\hat{\rho}_{n\mathbf{k}}$, respectively. Then, the wavefunctions of the spin-split bands are $|n\mathbf{k}\rangle \otimes |\uparrow; \hat{\rho}_{n\mathbf{k}}\rangle$ and $|n\mathbf{k}\rangle \otimes |\downarrow; \hat{\rho}_{n\mathbf{k}}\rangle$. Therefore, the spin is nearly fully polarized in each spin-split band, and the spin polarizations at Ga atoms and As atoms are parallel to each other.

We can further understand the direction of the spin polarization of each spin-split band. Since $\langle \mathbf{L}^{Ga} \rangle_{n\mathbf{k}}$ is anti-parallel to $\langle \mathbf{L}^{As} \rangle_{n\mathbf{k}}$ and both the orbital polarization and the atomic SOC of As are larger than those of Ga, $\hat{\rho}_{n\mathbf{k}}$ is parallel to $\langle \mathbf{L}^{As} \rangle_{n\mathbf{k}}$. Hence, the spin of the electronic states in the upper spin-split band, at both sublattices, aligns with $\langle \mathbf{L}^{As} \rangle_{n\mathbf{k}}$, and that in the lower spin-split band anti-aligns with $\langle \mathbf{L}^{As} \rangle_{n\mathbf{k}}$ (Figure 5). This behavior is different from the hidden spin polarization in centrosymmetric materials, in which the spin polarizations at the two sublattices are opposite to each other.

In addition, in GaAs or other non-centrosymmetric materials, if we decrease the strength of the SOC, the spin polarization of a spin-split band does not change appreciably because the eigenvectors of the full Hamiltonian are independent of the scaling of the spin-orbit interaction Hamiltonian in the small SOC limit. This behavior is different from the case of the hidden spin polarization in centrosymmetric materials, in which the magnitude scales linearly with the strength of the SOC in the same limit (Equation 1 and Figure 2).

Despite the fact that GaAs lacks inversion symmetry, its transport properties are effectively determined by the average of the spin-split bands depending on the level of impurity and temperature. For this reason, the *j*=3/2 Luttinger model[12] is commonly adopted in studying the transport properties of GaAs (e.g., see Reference 19). Although each spin-split band of GaAs is nearly fully spin polarized (Figures 5a-5d), when we average the spin polarization over the two spin-split bands, the spin polarization is very much reduced, but the orbital polarization is almost invariant upon averaging (Figures 5e and 5f). The averaged spin and orbital polarizations at As atoms (Figure 5f) are similar to the hidden spin and orbital polarizations at sublattice *A* in diamond (Figure 2b). In all cases, including diamond with a very large SOC of $\alpha^C = 1$ eV (Figure 2d), the band-averaged site-dependent spin polarization is much smaller in magnitude than the band-averaged site-dependent orbital polarization. These results indicate that site-dependent orbital polarizations are important in current-induced magnetization[5] of *both* centrosymmetric and non-centrosymmetric materials.

Recently, spin-polarized photocurrents were measured from bulk WSe$_2$,[20] a non-magnetic, centrosymmetric material. The results confirm the hidden spin polarization and the hidden orbital polarization, as the former is generated from the latter. Moreover, the hidden orbital polarization in materials with a small SOC can also be observed by measuring the spin-integrated photocurrents because it is not the spin polarization but the orbital polarization that determines the coupling between electrons and photons. Provided that the final state is well approximated by *s*-like states, the hidden orbital polarization also manifests itself in the circular dichroism of a non-magnetic, centrosymmetric material.

We now discuss the technological implications of our findings. When an electric current is applied to a centrosymmetric material, non-equilibrium, site-dependent orbital and spin magnetization can be generated. The current-induced magnetization is antiferromagnetic due to the nature of the hidden orbital and spin polarizations, and its direction depends on the direction of the current.[21] Antiferromagnetic spintronic devices, in which a current generates sublattice-dependent spin-orbit torques and changes the magnetic state of a material, have several advantages over conventional spintronic devices based on ferromagnetism. Since the total magnetic moment of an antiferromagnet is zero, antiferromagnetic devices are largely insensitive to the external environment and do not introduce magnetic crosstalk. Additionally, they operate much faster than ferromagnetic devices.[22]

The concept of hidden orbital polarization established here should be considered in properly predicting the site-dependent magnetism because, as we have shown, the spin polarization of a Bloch state could be much smaller than the orbital polarization in many materials (e.g., see Figure 2 and Figures 5e and 5f). Moreover, even in materials with weak SOC, the hidden orbital polarization can be used in antiferromagnetic information storage and processing because of the exchange interactions between localized, hidden orbital moments.[25]

To illustrate the idea that the current-induced hidden orbital polarization can play a more important role than the hidden spin polarization, we looked into the current-driven antiferromagnetism of silicon under a 2 % uniaxial compressive strain along the [001] direction, achievable in real experiments.[28,29] (Because silicon has many point group symmetries, an electric current in silicon does not generate site-dependent magnetization; however, a strain can result in current-induced magnetization by breaking some symmetries.[21,22]) Although silicon may not be the best material for antiferromagnetic information technology applications, it is one of the simplest and most well-known materials, a good candidate for supporting our hypothesis.

The effect of strain is simulated within our tight-binding model using Harrison's universal scaling method.[30] Following Reference 5, we obtain the non-equilibrium occupation factor $f'_{\nu\mathbf{k}}$ of a Bloch state $|\nu\mathbf{k}\rangle$ by considering the change from the equilibrium Fermi-Dirac occupation factor $f_{\nu\mathbf{k}} = f_{\text{FD}}(E_{\nu\mathbf{k}})$ of each Bloch state with the energy eigenvalue $E_{\nu\mathbf{k}}$:

$$f'_{\nu\mathbf{k}} = f_{\nu\mathbf{k}} + \frac{e\boldsymbol{\varepsilon}\tau}{\hbar} \cdot \frac{dE_{\nu\mathbf{k}}}{d\mathbf{k}} \frac{df_{\text{FD}}(E_{\nu\mathbf{k}})}{dE_{\nu\mathbf{k}}}$$

where $\tau$ denotes the scattering lifetime of charge carriers, $e$ the absolute value of the charge of an electron, and $\boldsymbol{\varepsilon}$ the applied electric field. The current-induced, site-dependent magnetization at the $A$ sublattice, $\mathbf{M}^A$, is then given by

$$\mathbf{M}^A = -\frac{\mu_B}{\hbar} \sum_\nu \int_{\text{BZ}} \frac{d^3k}{(2\pi)^3} f'_{\nu\mathbf{k}} (\langle \mathbf{L}^A \rangle_{\nu\mathbf{k}} + 2\langle \mathbf{S}^A \rangle_{\nu\mathbf{k}}),$$

where $\mu_B$ is the Bohr magneton. As in Reference 5, we assumed that the spin g-factor of electrons is 2.

Figure 6 shows the calculated contributions of the orbital and spin polarizations to the induced magnetization of strained, hole-doped silicon at sublattice $A$ per unit strength of the electric field as a function of the doping concentration $n_p$. The scattering lifetime $\tau$ at each $n_p$ is extracted from the measured mobility data[31] by using the Drude model.

Clearly, the orbital contribution to the current-induced antiferromagnetism is much larger than the spin contribution. Additionally, the induced magnetization at each site of silicon can be larger than the total induced magnetization of $Cr_2O_3$, the most well-known magnetoelectric material, with a $\mu_0 dM/d\varepsilon$ value of approximately 1 ps/m.[32] Again, we are not claiming that compressed silicon is the best material for antiferromagnetic information technology exploiting the hidden orbital polarization; larger current-induced antiferromagnetism is expected in lower-symmetry materials. However, our proof-of-concept calculations illustrate that it is a worthwhile research direction to search for materials with large hidden orbital polarizations useful in antiferromagnetic information technology, irrespective of the size of the spin-orbit coupling. This result shows that investigating the effect of the hidden orbital polarization on antiferromagnetic information storage and processing is an important and promising theoretical and experimental future research direction.

**Figure 6.** (a) The orbital and spin contributions to the current-induced, site-dependent magnetization of silicon under a 2 % uniaxial compressive strain along [001] versus the hole concentration $n_p$. (b) The same quantity divided by the scattering lifetime of holes.

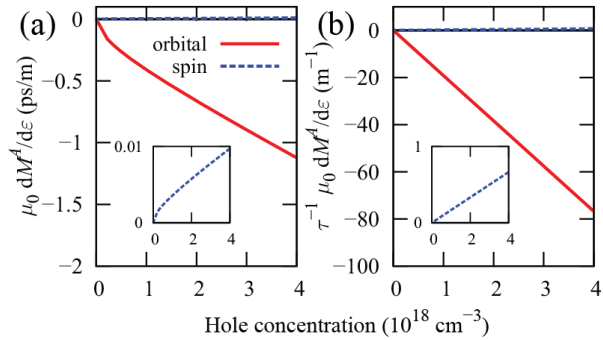

## CONCLUSIONS

In conclusion, we have shown that even in centrosymmetric, non-magnetic materials, there can exist large site-dependent, hidden orbital polarizations. In centrosymmetric group IV materials such as diamond, Si, and Ge, the hidden spin polarization is very small, but the hidden orbital polarization is on the order of $\hbar$. We have also found, using a general perturbative scheme that is applicable not only to diamond, Si, and Ge (with small hidden spin polarizations) but also to layered materials such as $MoS_2$ and $WSe_2$ with hidden spin polarizations close to the maximum value, that the hidden spin polarization is completely determined by the site-dependent orbital angular momentum in general centrosymmetric, non-magnetic materials. If the energy distance between nearby bands is comparable to or smaller than the atomic spin orbit coupling, the hidden spin polarization is large. In the case of zinc-blende materials, this energy difference (nearest-neighbor hopping) is a few eV, and in the case of transition-metal dichalcogenides, this energy difference (interlayer hopping) is a few tens of meV. In any case, however, first-order or higher-order perturbative theory with respect to the SOC connects the hidden spin polarization to site-dependent orbital angular momenta. By comparing the strength of the SOC and the interlayer hopping constant in $MoS_2$ and $WSe_2$, we have shown that the hidden spin polarization in transition metal dichalcogenides can be significantly reduced by applying a pressure. Our study also illustrates that site-dependent orbital polarizations play an important role in current-induced magnetization of both centrosymmetric materials and non-centrosymmetric materials such as GaAs. We have discussed the experimental signatures of the hidden orbital polarization in centrosymmetric materials in both spin-resolved and -integrated photoemission spectroscopies. We have also calculated the current-driven antiferromagnetism in compressed silicon and have shown that an appreciable amount of orbital (antiferro-)magnetization can be induced even when the spin counterpart is negligible, demonstrating the potentially important role of hidden orbital polarizations in antiferromagnetic information technology. Because there are more degrees of freedom in orbital polarization than in spin polarization, the hidden orbital polarization may lead to new discoveries in physics.

## CONFLICT OF INTEREST

The authors declare no conflict of interest.

## ACKNOWLEDGMENTS

We thank Tae Yun Kim for discussions at an early stage of this work and for drawing our attention to Reference 22 and Ivo Souza for discussions on many aspects of the orbital magnetization of solids and for pointing out the required symmetry lowering in the current-induced magnetization.[21] This work was supported by the Creative-Pioneering Research Program through Seoul National University.

[1]     Zhang, X., Liu, Q., Luo, J.-W., Freeman, A. J. & Zunger, A. Hidden spin polarization in inversion-symmetric bulk crystals. *Nat. Phys.* **10**, 387 (2014).

[2]     Reck, R. A. & Fry, D. L. Orbital and spin magnetization in Fe-Co, Fe-Ni, and Ni-Co. *Phys. Rev.* **184**, 492 (1969).

[3]     Ceresoli, D., Gerstmann, U., Seitsonen, A. P. & Mauri, F. First-principles theory of orbital magnetization. *Phys. Rev. B* **81**, 060409 (2010).

[4]     Taylor, J. W., Duffy, J. A., Bebb, A. M., Lees, M. R., Bouchenoire, L., Brown, S. D. & Cooper, M. J. Temperature dependence of the spin and orbital magnetization density in $Sm_{0.982}Gd_{0.018}Al_2$ around the spin-orbital compensation point. *Phys. Rev. B* **66**, 161319 (2002).

[5]     Yoda, T., Yokoyama, T. & Murakami, S. Current-induced orbital and spin magnetizations in crystals with helical structure. *Sci. Rep.* **5**, 12024 (2015).

[6]     Zhong, S., Moore, J. E. & Souza, I. Gyrotropic magnetic effect and the magnetic moment on the Fermi surface. *Phys. Rev. Lett.* **116**, 077201 (2016).

[7]     Park, S. R., Kim, C. H., Yu, J., Han, J. H. & Kim, C. Orbital-angular-momentum based origin of Rashba-type surface band splitting. *Phys. Rev. Lett.* **107**, 156803 (2011).

[8]     Park, J.-H., Kim, C. H., Rhim, J.-W. & Han, J. H. Orbital Rashba effect and its detection by circular dichroism angle-resolved photoemission spectroscopy. *Phys. Rev. B* **85**, 195401 (2012).

[9] Kim, B., Kim, C. H., Kim, P., Jung, W., Kim, Y., Koh, Y., Arita, M., Shimada, K., Namatame, H., Taniguchi, M., Yu, J. & Kim, C. Spin and orbital angular momentum structure of Cu(111) and Au(111) surface states. *Phys. Rev. B* **85**, 195402 (2012).

[10] Baidya, S. Waghmare, U. V., Paramekanti, A., Saha-Dasgupta, T. High-temperature large-gap quantum anomalous Hall insulating state in ultrathin double perovskite films. *Phys. Rev. B* **94**, 155405 (2016).

[11] Vogl, P., Hjalmarson, H. P. & Dow, J. D. A semi-empirical tight-binding theory of the electronic structure of semiconductors. *J. Phys. Chem. Solids* **44**, 365 (1983).

[12] Luttinger, J. M. Quantum theory of cyclotron resonance in semiconductors: general theory. *Phys. Rev.* **102**, 1030 (1956).

[13] Petersen, L. & Hedegård, P. A simple tight-binding model of spin–orbit splitting of *sp*-derived surface states. *Surf. Sci.* **459**, 49 (2000).

[14] Waugh, J. A., Nummy, T., Parham, S., Liu, Q., Zhang, X., Zunger, A. & Dessau, D. S. Minimal ingredients for orbital texture switches at Dirac points in strong spin-orbit coupled materials. *arXiv*:1608.01387v1 (2016).

[15] We can use the known matrix elements of $L_x = (L_+ + L_-)/2$, $L_y = (L_+ - L_-)/2i$ and $L_z$ in $|lm\rangle$ basis. These $|lm\rangle$ basis functions can be expressed using cubic harmonics; for example, $|l = 2, m = \pm 2\rangle = (|d_{x^2-y^2}\rangle \pm i|d_{xy}\rangle)/\sqrt{2}$, $|l = 2, m = \pm 1\rangle = (\mp|d_{xz}\rangle - i|d_{yz}\rangle)/\sqrt{2}$, and $|l = 2, m = 0\rangle = |d_{z^2}\rangle$.

[16] Fang, S., Kuate Defo, R., Shirodkar, S. N., Lieu, S., Tritsaris, G. A. & Kaxiras, E. *Ab initio* tight-binding Hamiltonian for transition metal dichalcogenides. *Phys. Rev. B* **92**, 205108 (2015).

[17] Zhao, Z., Zhang, H., Yuan, H., Wang, S., Lin, Y., Zeng, Q., Xu, G., Liu, Z., Solanki, G.K., Patel, K.D. and Cui, Y., Hwang, H. Y. & Mao, W. L. Pressure induced metallization with absence of structural transition in layered molybdenum diselenide. *Nat. Commun.* **6**, 7312 (2015).

[18]  Chadi, D. J. Spin-orbit splitting in crystalline and compositionally disordered semiconductors. *Phys. Rev. B* **16**, 790 (1977).

[19]  Murakami, S., Nagaosa, N. & Zhang, S.-C. SU(2) non-Abelian holonomy and dissipationless spin current in semiconductors. *Phys. Rev. B* **69**, 235206 (2004).

[20]  Riley, J. M., Mazzola, F., Dendzik, M., Michiardi, M., Takayama, T., Bawden, L., Granerød, C., Leandersson, M., Balasubramanian, T., Hoesch, M., Kim, T. K., Takagi, H., Meevasana, W., Hofmann, P., Bahramy, M. S., Wells, J. W. & King, P. D. C. Direct observation of spin-polarized bulk bands in an inversion-symmetric semiconductor. *Nat. Phys.* **10**, 835 (2014).

[21]  We note that a symmetry lowering, for example by strain, is required for diamond, Si, Ge, and GaAs to generate a current-induced, site-dependent magnetization[22]; on the other hand, there are other classes of materials that do not require an additional symmetry lowering including $CuMnAs$[23] or $Mn_2Au$[24], which are also intrinsically antiferromagnetic.

[22]  Jungwirth, T., Marti, X., Wadley, P. & Wunderlich, J. Antiferromagnetic spintronics. *Nat. Nanotechnol.* **11**, 231 (2016).

[23]  Wadley, P., Howells, B., Železný, J., Andrews, C., Hills, V., Campion, R. P., Novák, V., Olejník, K., Maccherozzi, F., Dhesi, S. S., Martin, S. Y., Wagner, T., Wunderlich, J., Freimuth, F., Mokrousov, Y., Kuneš, J., Chauhan, J. S., Grzybowski, M. J., Rushforth, A. W., Edmonds, K. W., Gallagher, B. L. & Jungwirth, T. Electrical switching of an antiferromagnet. *Science* **351**, 587 (2016).

[24]  Železný, J., Gao, H., Výborný, K., Zemen, J., Mašek, J., Manchon, A., Wunderlich, J., Sinova, J. & Jungwirth, T. Relativistic Néel-Order fields induced by electrical current in antiferromagnets. *Phys. Rev. Lett.* **113**, 157201 (2014).

[25]  Details on the exchange interactions between orbital moments can be found in, e.g., Reference 26 and 27.

[26]     Kugel', K. I., Khomskii, D. I. The Jahn-Teller effect and magnetism: transition metal compounds. *Sov. Phys. Usp.* **25**, 231 (1982).

[27]     Khaliullin, G. Orbital Order and Fluctuations in Mott Insulators. *Prog. Theor. Phys. Supp.* **160**, 155 (2005).

[28]     Laude, L. D., Pollak, F. H. & Cardona, M. Effects of Uniaxial Stress on the Indirect Exciton Spectrum of Silicon. *Phys. Rev. B* **3**, 2623 (1971).

[29]     Milne, J. S., Favorskiy, I., Rowe, A. C. H., Arscott, S., & Renner, C. Piezoresistance in silicon at uniaxial compressive stresses up to 3 GPa. *Phys. Rev. Lett.* **108**, 256801 (2012).

[30]     Harrison, W. A. *Electronic Structure and the properties of Solids* (Freeman, San Francisco, USA, 1980).

[31]     Hu, C. *Modern Semiconductor Devices for Integrated Circuits* (Prentice Hall, New Jersey, USA, 2010).

[32]     Wiegelmann, H., Jansen, A. G. M., Wyder, P., Rivera, J. P., & Schmid, H. Magnetoelectric effect of $Cr_2O_3$ in strong static magnetic fields. *Ferroelectrics* **162**, 141 (1994)

# Hidden orbital polarization in diamond, silicon, germanium, gallium arsenide and layered materials: Supplementary information

Ji Hoon Ryoo and Cheol-Hwan Park*

Department of Physics, Seoul National University, Seoul 08826, Korea

E-mail: cheolhwan@snu.ac.kr

**Supplementary Discussion 1: Site-dependent spin and orbital polarizations of the highest-energy valence bands of diamond, Si, Ge, and GaAs**

**Figure S1.** The local orbital and spin textures of the two highest-energy (double-degenerate) valence bands of diamond, Si and Ge on $k_z = 0$ plane.

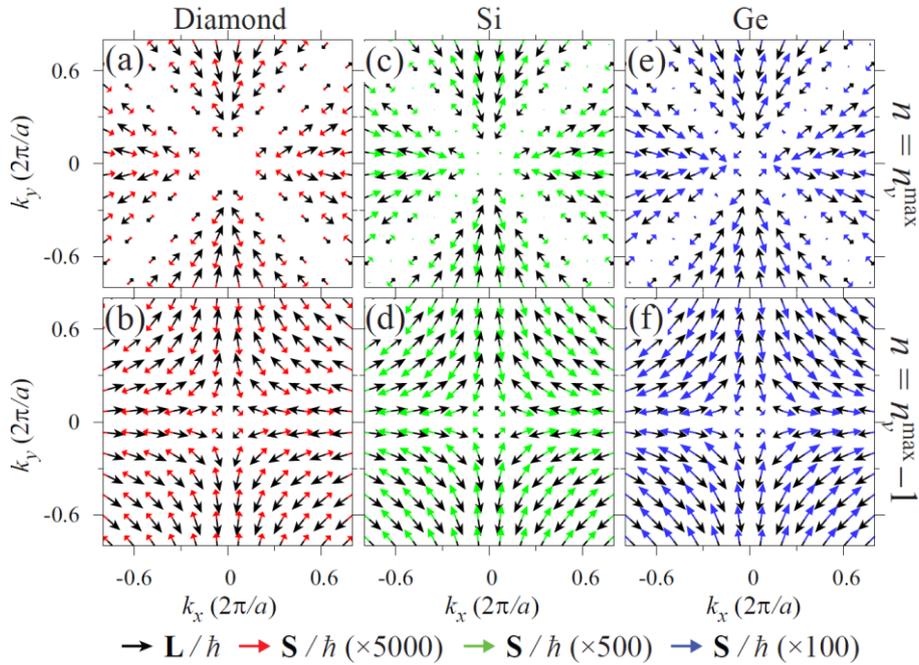

**Figure S2**. The site-dependent orbital and spin textures of the highest-energy valence bands of GaAs on $k_z = 0$ plane. (a) and (b), (c) and (d), and (e) and (f) show the local spin and orbital polarizations of the upper spin-split band, those of the lower spin-split band, and their averages, respectively.

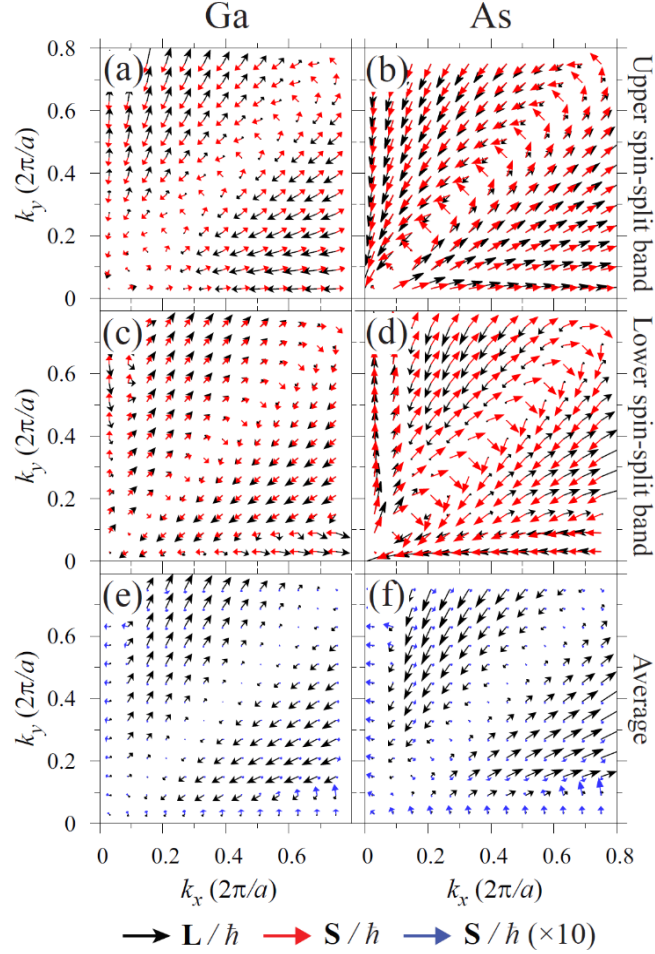

**Supplementary Discussion 2: Analytic expression of the magnitude of hidden spin polarization of MoS$_2$ as a function of the strength of SOC and the interlayer coupling**

In this section, we present a theory explaining the hidden spin polarization of the highest-energy valence bands at K of MoS$_2$, which includes higher-order perturbations not considered in Equation 1 of the main paper. First, let us neglect both SOC and interlayer hopping and denote the orbital part of the wavefunction at the valence band maximum at K of the upper layer as $|\psi^{(0)}_{K,\text{upper}}\rangle$ and that of the lower layer as $|\psi^{(0)}_{K,\text{lower}}\rangle$. The spin-orbit coupling $\alpha(K)^{\text{MoS}_2}$ and the interlayer coupling $t(\text{K})_{\text{inter}}^{\text{MoS}_2}$ of MoS$_2$ defined in the previous section can be redefined as follows (we show later that the two definitions are consistent):

$$\alpha(\text{K})^{\text{MoS}_2} = 2\left\langle \psi_{\text{K,lower}}^{(0)} \uparrow \middle| \Delta H_{\text{SOC}} \middle| \psi_{\text{K,lower}}^{(0)} \uparrow \right\rangle = 2\left\langle \psi_{\text{K,upper}}^{(0)} \downarrow \middle| \Delta H_{\text{SOC}} \middle| \psi_{\text{K,upper}}^{(0)} \downarrow \right\rangle$$

$$= -2\left\langle \psi_{\text{K,upper}}^{(0)} \uparrow \middle| \Delta H_{\text{SOC}} \middle| \psi_{\text{K,upper}}^{(0)} \uparrow \right\rangle = -2\left\langle \psi_{\text{K,lower}}^{(0)} \downarrow \middle| \Delta H_{\text{SOC}} \middle| \psi_{\text{K,lower}}^{(0)} \downarrow \right\rangle,$$

where in the first line we have used the inversion and time reversal symmetries and in the second line we have used the symmetry operation which acts as the identity operator in the orbital space and the time reversal operator in the spin space, and

$$t(\text{K})_{\text{inter}}^{\text{MoS}_2} = \left\langle \psi_{\text{K,upper}}^{(0)} \middle| \Delta H_{\text{inter}} \middle| \psi_{\text{K,lower}}^{(0)} \right\rangle.$$

Here, $\left| \psi_{\text{K,upper}}^{(0)} \uparrow \right\rangle = \left| \psi_{\text{K,upper}}^{(0)} \right\rangle \otimes |\uparrow\rangle$ and so on, and $\Delta H_{\text{SOC}} = \sum_j \alpha^j \mathbf{L}^j \cdot \mathbf{S}^j / \hbar^2$ is the contribution of the on-site spin-orbit coupling to the Hamiltonian. ($\alpha^j$ is the strength of atomic spin-orbit coupling of an atom $j$, and $\mathbf{L}^j$ and $\mathbf{S}^j$ are the orbital and spin angular momentum operators projected on an atom $j$, respectively.) $\Delta H_{\text{inter}}$ is the contribution to the Hamiltonian of the interlayer hopping between the outer-most S atoms. Interlayer hopping integrals are modeled using Slater-Koster type parameters in [S. Fang et al., Phys. Rev. B **92**, 205108 (2015)].

If the weak interlayer hopping is added to the Hamiltonian (but SOC is still neglected), the eigenstates are now given by $|\pm, \text{K}\rangle = \left( \left| \psi_{\text{K,upper}}^{(0)} \right\rangle \pm \left| \psi_{\text{K,lower}}^{(0)} \right\rangle \right)/\sqrt{2}$ and the energy difference between these two states is $E_{+,\text{K}} - E_{-,\text{K}} = 2t(\text{K})_{\text{inter}}^{\text{MoS}_2}$. Due to the absence of SOC so far, bands are still spin-degenerate and hence the hidden spin polarizations are zero.

Now let us turn very weak SOC on. The natural generalization of Equation 1 in the main paper to centrosymmetric crystals where each sublattice comprising the crystal have multiple atoms is

$$\langle \mathbf{S}^\beta \rangle_{n\mathbf{k}}^{\text{avg}} = \frac{1}{2} \sum_{m \neq n} \frac{\langle n\mathbf{k} | P^\beta | m\mathbf{k} \rangle \langle m\mathbf{k} | \sum_j \alpha^j \mathbf{L}^{\beta,j} | n\mathbf{k} \rangle + \text{c.c.}}{E_{n\mathbf{k}} - E_{m\mathbf{k}}},$$

where $n$ and $m$ are band indices, $j$ is the atomic index in a given sublattice and $\beta$ the sublattice index. (In MoS$_2$, $\beta$ runs over the upper and lower layers and $j$ runs over an Mo atom and two S atoms.) Applying this equation to $\left| \psi_{\text{K},+}^{(0)} \right\rangle$ in MoS$_2$, we get

$$\langle \mathbf{S}^{\text{lower}} \rangle_{+,\text{K}}^{\text{avg}} = \frac{1}{2} \sum_{m \neq +} \frac{\langle +, \text{K} | P^{\text{lower}} | m, \text{K} \rangle \langle m, \text{K} | \sum_j \alpha^j \mathbf{L}^{\text{lower},j} | +, \text{K} \rangle + \text{c.c.}}{E_{+,\text{K}} - E_{m,\text{K}}}$$

$$\approx \frac{1}{2} \frac{\langle +, \text{K} | P^{\text{lower}} | -, \text{K} \rangle \langle -, \text{K} | \sum_j \alpha^j \mathbf{L}^{\text{lower},j} | +, \text{K} \rangle + \text{c.c.}}{E_{+,\text{K}} - E_{-,\text{K}}}.$$

In the last line, we used the fact that the energy difference between $|+, \text{K}\rangle$ and other bands except $|-, \text{K}\rangle$ is much larger than $E_{+,\text{K}} - E_{-,K} = 2t(\text{K})_{\text{inter}}^{\text{MoS}_2}$ because the interlayer hopping is much weaker than the intralayer hopping, and hence considered only the contribution of $|-, \text{K}\rangle$. Using $\langle +, \text{K} | P^{\text{lower}} | -, \text{K} \rangle = -1/2$ and

$$\sum_j \langle -, K | \alpha^j \mathbf{L}^{\text{lower},j} | +, K \rangle = \left(-\frac{1}{2}\right) \sum_j \langle \psi^{(0)}_{K,\text{lower}} | \alpha^j \mathbf{L}^{\text{lower},j} | \psi^{(0)}_{K,\text{lower}} \rangle$$

$$= \left(-\frac{\hat{z}}{2}\right) \sum_j \langle \psi^{(0)}_{K,\text{lower}} | \alpha^j L_z^{\text{lower},j} | \psi^{(0)}_{K,\text{lower}} \rangle$$

$$= \left(-\frac{\hat{z}}{\hbar}\right) \sum_j \langle \psi^{(0)}_{K,\text{lower}} \uparrow | \alpha^j L_z^{\text{lower},j} S_z^{\text{lower},j} | \psi^{(0)}_{K,\text{lower}} \uparrow \rangle = -\hbar\, \alpha(K)^{\text{MoS}_2} \hat{z},$$

where in the second line the $x$ and $y$ components of the hidden orbital polarizations at K vanish because of the three-fold rotational symmetry of the crystal, we get

$$\langle \mathbf{S}^{\text{lower}} \rangle^{\text{avg}}_{+,K} \approx \frac{\hbar}{4} \frac{\alpha(K)^{\text{MoS}_2}}{t(K)^{\text{MoS}_2}_{\text{inter}}} \hat{z}.$$

Thus, the hidden spin polarization in MoS$_2$ at K is proportional to the strength of SOC when SOC is weak.

We can generalize the perturbation theory further to make it applicable regardless of whether SOC is weaker than the interlayer coupling or not, as long as both the strength of SOC and the interlayer coupling are weaker than the *intra*-layer coupling. Projecting the perturbations to the Hamiltonian into the four-dimensional Hilbert space spanned by $|\psi^{(0)}_{K,\text{upper}} \uparrow\rangle$, $|\psi^{(0)}_{K,\text{upper}} \downarrow\rangle$, $|\psi^{(0)}_{K,\text{lower}} \uparrow\rangle$, and $|\psi^{(0)}_{K,\text{lower}} \downarrow\rangle$, we get the following 4 by 4 effective Hamiltonian:

$$H_{\text{eff}} = \begin{pmatrix} -\alpha(K)^{\text{MoS}_2} & 0 & t(K)^{\text{MoS}_2}_{\text{inter}} & 0 \\ 0 & \alpha(K)^{\text{MoS}_2} & 0 & t(K)^{\text{MoS}_2}_{\text{inter}} \\ t(K)^{\text{MoS}_2}_{\text{inter}} & 0 & \alpha(K)^{\text{MoS}_2} & 0 \\ 0 & t(K)^{\text{MoS}_2}_{\text{inter}} & 0 & -\alpha(K)^{\text{MoS}_2} \end{pmatrix},$$

where we have used the definitions of $\alpha(K)^{\text{MoS}_2}$ and $t(K)^{\text{MoS}_2}_{\text{inter}}$ presented in the beginning of this section. Note that since the orbital polarization of a monolayer is along $z$, spin-up states and spin-down states do not intermix in the leading order. Solving this effective Hamiltonian for the eigenstates $|\pm', K\rangle \otimes |\uparrow\rangle$ and $|\pm', K\rangle \otimes |\downarrow\rangle$, we get

$$\langle \mathbf{S}^{\text{lower}} \rangle^{\text{avg}}_{+',K} = \frac{\langle \mathbf{S}^{\text{lower}} \rangle_{+',K,\uparrow} + \langle \mathbf{S}^{\text{lower}} \rangle_{+',K,\downarrow}}{2} = \frac{\hbar}{4} \frac{\alpha(K)^{\text{MoS}_2}}{\sqrt{\alpha(K)^{\text{MoS}_2\,2} + t(K)^{\text{MoS}_2\,2}_{\text{inter}}}} \hat{z},$$

which explains the saturation of the hidden spin polarization when SOC is stronger than the interlayer coupling as well as the linear dependence of the hidden spin polarization on the strength of SOC when SOC is weaker than the interlayer coupling.

**Supplementary Discussion 3: Group-theoretical analysis on hidden orbital polarization**

Regarding the relation between the hidden orbital polarization and bulk / site symmetries of a crystal, since both the spin and angular momenta are pseudo-vectors which are odd under time reversal, the classification scheme in Reference 1 of the main paper should be equally applicable to the case of hidden orbital polarizations: the only necessary condition for the hidden orbital and spin polarizations to be nonzero is that the inversion center does not lie at an atomic position. Reference 1 of the main paper further classifies centrosymmetric materials satisfying such a condition into R-2 and D-2 types, according to whether the site symmetry group of any atom allows the existence of a nonzero polar vector.

On the other hand, although the condition to have a finite hidden orbital polarization is the same as the condition to have a finite hidden spin polarization, i.e., the inversion centers not lying at atomic positions, we note that it is much easier to find large hidden orbital polarizations than large spin polarizations in nature (see Figure 3 and 4 in the main manuscript and Figure S3). To have a large hidden spin polarization, we need an additional condition to be satisfied that the energy separation from nearby bands be smaller than the strength of SOC as we have shown in the previous section. Therefore, hidden orbital polarizations are far more abundant in nature than hidden spin polarizations.

**Figure S3.** (a) The atomic structure of bulk BN and (b) the lower-layer hidden polarizations along $z$ of the highest-energy valence bands among $sp^2$ valence bands at K as a function of the strength of SOC. The fully polarized values are $\langle S^{\text{lower}} \rangle_{\text{avg}}^{\text{max}} = \frac{\hbar}{2} \times 0.5$ and $\langle L^{\text{lower}} \rangle_{\text{avg}}^{\text{max}} = \hbar \times 0.5$. The hopping integrals are taken from [J. Widany, Density Functional Tight Binding Calculations on the Structure of Complex Boron Nitride Systems, PhD dissertation, Technische Universität Berlin (1997)]. For demonstration purposes, we set the strengths of SOC of boron and nitrogen atoms to be equal. The physical values of the atomic SOC of boron and nitrogen atoms are 3~6 meV, i.e., almost zero in the scale of the figure.

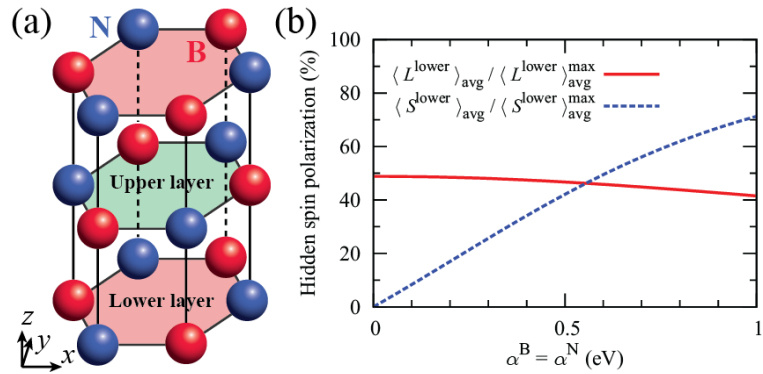

We find that once a centrosymmetric crystal does not have an inversion center at atomic positions, in principle, it can have large hidden spin and orbital polarizations, regardless of whether a material belongs to R-2 or D-2 class. We have shown in Figure 3 and 4 in the main manuscript that R-2 materials such as $MoS_2$ or $WSe_2$ can have large hidden spin and orbital polarizations. Similarly, Figure S3 shows that a system of D-2 type can, in principle, have large hidden spin and orbital polarizations, i.e., no symmetry principle prevents D-2 type materials from having large hidden polarizations.

**Supplementary Discussion 4: Anti-alignment of the hidden orbital and spin polarizations**

In the case where we consider only the nearest-neighbor hopping between the *p* orbitals in diamond and neglect the mixing between *s* and *p* orbitals, we can understand the (anti-)alignment of the hidden orbital and spin polarization as follows.

Since the last line of Equation 1 in the main paper does not involve a spin index, we can interpret the hidden spin polarization in another way: Suppose that $|n\mathbf{k}\rangle$ is the orbital part of an eigenstate of the Hamiltonian without SOC. Then, up to first order in SOC, the site-dependent spin polarization $\langle \mathbf{S}^A \rangle_{n\mathbf{k}}^{\text{avg}}$ that $|n\mathbf{k}\rangle$ acquires is proportional to the change in site-dependent orbital polarization $\langle \mathbf{L}^A \rangle_{n\mathbf{k}}$ in the system without SOC when perturbed by $P^A$, an inversion-symmetry-breaking on-site potential.

If we neglect *s* orbitals and focus on *p* orbitals and the nearest-neighbor hopping, the Hamiltonian satisfies $UH(\mathbf{k})U^{-1} = -H(\mathbf{k})$ for any $\mathbf{k}$ (we set the on-site energy of the *p* orbitals to zero), where $U = P^A - P^{\bar{A}}$ is a unitary operator (recall that $P^\beta$ is the projection operator onto $\beta = A, \bar{A}$ sublattices). For any $\mathbf{k}$ where the band does not cross the on-site energy of the *p* orbitals, let the orbital part of a state in the valence band be $|n\mathbf{k}\rangle = \sum_{i=x,y,z}(c_i^{n\mathbf{k}}|p_i, A; \mathbf{k}\rangle + d_i^{n\mathbf{k}}|p_i, \bar{A}; \mathbf{k}\rangle)$. The corresponding state in the conduction band is $U|n\mathbf{k}\rangle = \sum_{i=x,y,z}(c_i^{n\mathbf{k}}|p_i, A; \mathbf{k}\rangle - d_i^{n\mathbf{k}}|p_i, \bar{A}; \mathbf{k}\rangle)$ and $(\mathbf{c}^{n\mathbf{k}})^* \cdot \mathbf{c}^{m\mathbf{k}} = (\mathbf{d}^{n\mathbf{k}})^* \cdot \mathbf{d}^{m\mathbf{k}} = \delta_{nm}/2$ ($n, m = 1,2,3$).

Now, suppose we perturb this system by $P^A$ operator. By the orthogonality of $\{\mathbf{c}^{n\mathbf{k}}|n = 1,2,3\}$ and that of $\{\mathbf{d}^{n\mathbf{k}}|n = 1,2,3\}$, $P^A$ mixes $|n\mathbf{k}\rangle$ only with $U|n\mathbf{k}\rangle$ and not with other four states (i.e. $|m\mathbf{k}\rangle$ and $U|m\mathbf{k}\rangle$ with $m \neq n$). Therefore, the first-order correction to $|n\mathbf{k}\rangle$ by the perturbation, $\Delta|n\mathbf{k}\rangle$, is proportional to $U|n\mathbf{k}\rangle$. Since the ratio between the complex amplitudes of $|p_x, A; \mathbf{k}\rangle$, $|p_y, A; \mathbf{k}\rangle$, and $|p_z, A; \mathbf{k}\rangle$ for the new eigenvector $|n\mathbf{k}\rangle + \Delta|n\mathbf{k}\rangle$ is still $c_x^{n\mathbf{k}}: c_y^{n\mathbf{k}}: c_z^{n\mathbf{k}}$, the same as without perturbation, $\langle \mathbf{L}^A \rangle_{n\mathbf{k}}$ does not change its direction. Going back to the original problem, we conclude that $\langle \mathbf{S}^A \rangle_{n\mathbf{k}}^{\text{avg}}$ acquired by small SOC is parallel or anti-parallel to the orbital polarization.

When we consider both s and p orbitals and the nearest-neighbor hopping among them, it is still true that the spin polarization is exactly (anti-)parallel to the orbital polarization in the limit of zero spin-orbit coupling, according to the numerical calculation, although we were not able to find an analytic proof. However, if we add next-nearest-neighbor hopping to our model, the angle between $\langle \mathbf{L}^A \rangle_{n\mathbf{k}}^{\text{avg}}$ and $\langle \mathbf{S}^A \rangle_{n\mathbf{k}}^{\text{avg}}$ decreases from 180°.

An interesting scenario is that the anti-alignment between the hidden spin and orbital polarizations could be explained using some form of an energy-minimization argument (e.g., minimizing the $\mathbf{L} \cdot \mathbf{S}$ term). The energy-minimization argument holds for materials without inversion symmetry, where the degeneracy in energy is broken to first order in SOC, since the minimization of the (quasiparticle, not the total) energy in the spin space is mathematically equivalent to the diagonalization of the two-by-two matrix generated by projecting the Hamiltonian to the degenerate states which have the same orbital part of the wavefunction. For example, suppose that the orbital part of a Bloch state of GaAs obtained without considering SOC is $|n\mathbf{k}\rangle$ and the two spin-degenerate basis states are $|n\mathbf{k}\rangle \otimes |\uparrow\rangle$ and $|n\mathbf{k}\rangle \otimes |\downarrow\rangle$. Then, SOC splits the two states and makes the spin polarization parallel (or antiparallel) to $\langle n\mathbf{k}|[\alpha^{\text{Ga}}\mathbf{L}^{\text{Ga}} + \alpha^{\text{As}}\mathbf{L}^{\text{As}}]|n\mathbf{k}\rangle$, as explained in the main paper.

However, in the case of centrosymmetric materials, this energy-minimization argument is not justified. Consider the orbital part $|n\mathbf{k}\rangle$ of a Bloch state of, e.g., diamond. The matrix element $\langle n\mathbf{k}|[\alpha^C\mathbf{L}^A + \alpha^C\mathbf{L}^{\bar{A}}]|n\mathbf{k}\rangle$ vanishes in this case and the determination of hidden spin polarization necessarily involves the interband matrix elements of the (site-dependent) orbital angular momentum operator, as shown in Equation 1 of the manuscript. Also, as mentioned above, when the next-nearest-neighbor hopping is considered, the angle between the hidden spin and orbital polarizations deviates from 180 degrees (even in the limit SOC approaches zero).

We further notice that whether the hidden spin and orbital polarizations are antiparallel or parallel to each other is not determined by the band index alone but it is also dependent on the Bloch wavevector as Figure S4 shows. Note that except at Γ there is no band crossing in the momentum-space shown in Figure S4 and hence the **k**-dependent parallel / antiparallel alignment of the hidden spin and orbital polarizations occurs in the (doubly-degenerate) lowest conduction bands, which are well-separated from other energy bands.

**Figure S4.** The average site-dependent spin-orbital texture ($k_z = 0$) at sublattice $A$ of the doubly-degenerate, lowest-energy conduction bands of diamond.

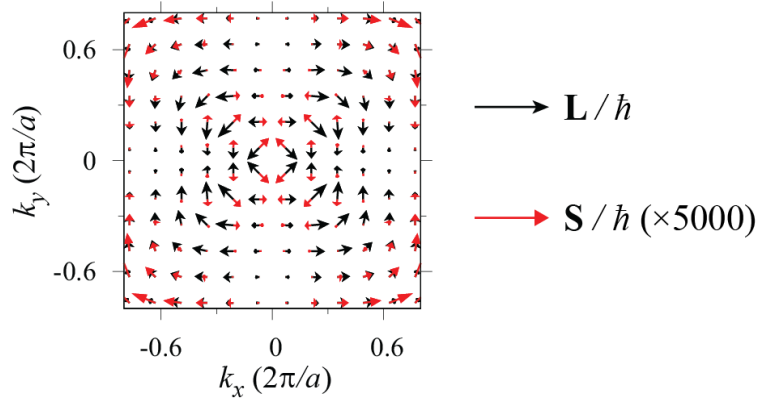